\def\year{2021}\relax
\documentclass[letterpaper]{article} 
\usepackage{aaai21}  
\usepackage{times}  
\usepackage{helvet} 
\usepackage{courier}  
\usepackage[hyphens]{url}  
\usepackage{graphicx} 
\usepackage{graphicx}  
\usepackage{natbib}  
\usepackage{caption} 
\usepackage{subcaption}
\usepackage{amsmath}
\DeclareMathOperator*{\argmax}{argmax} 
\usepackage{amssymb}
\usepackage{multirow}
\usepackage{romannum}

\title{Non-invasive Self-attention for Side Information Fusion in Sequential Recommendation}
\author {

        Chang Liu\textsuperscript{\rm 1},
        Xiaoguang Li\textsuperscript{\rm 2}\thanks{Equal contribution}, 
        Guohao Cai\textsuperscript{\rm 2},  
        Zhenhua Dong\textsuperscript{\rm 2}, 
        Hong Zhu\textsuperscript{\rm 2}, 
        Lifeng Shang\textsuperscript{\rm 2}  \\
}
\affiliations {
    \textsuperscript{\rm 1} The University of Hong Kong,  
    \textsuperscript{\rm 2} Huawei Noah's Ark Lab \\
    lcon7@connect.hku.hk, \{lixiaoguang11, caiguohao1, dongzhenhua, zhuhong8, shang.lifeng\}@huawei.com
}

\usepackage[switch]{lineno}
\begin{document}

\maketitle

\begin{abstract}
Sequential recommender systems aim to model users’ evolving interests from their historical behaviors, and hence make customized time-relevant recommendations. Compared with traditional models, deep learning approaches such as CNN and RNN have achieved remarkable advancements in recommendation tasks. Recently, the BERT framework also emerges as a promising method, benefited from its self-attention mechanism in processing sequential data. However, one limitation of the original BERT framework is that it only considers one input source of the natural language tokens. It is still an open question to leverage various types of information under the BERT framework. Nonetheless, it is intuitively appealing to utilize other side information, such as item category or tag, for more comprehensive depictions and better recommendations. In our pilot experiments, we found naive approaches, which directly fuse types of side information into the item embeddings, usually bring very little or even negative effects. Therefore, in this paper, we propose the \textbf{NO}n-in\textbf{V}asive self-\textbf{A}ttention mechanism (NOVA) to leverage side information effectively under the BERT framework. NOVA makes use of side information to generate better attention distribution, rather than directly altering the item embeddings, which may cause information overwhelming. We validate the NOVA-BERT model on both public and commercial datasets, and our method can stably outperform the state-of-the-art models with negligible computational overheads.
\end{abstract}

\begin{figure}[ht]
  \centering
  \begin{subfigure}{\columnwidth}
    \centering
    \includegraphics[scale=0.5]{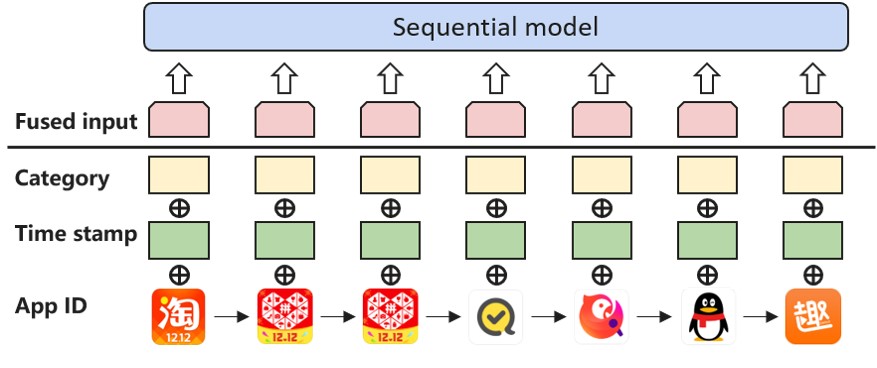}
    \caption{Invasive fusion approaches}
    \label{fig:intro-invasive}
  \end{subfigure}
  
  \begin{subfigure}{\columnwidth}
    \centering
    \includegraphics[scale=0.5]{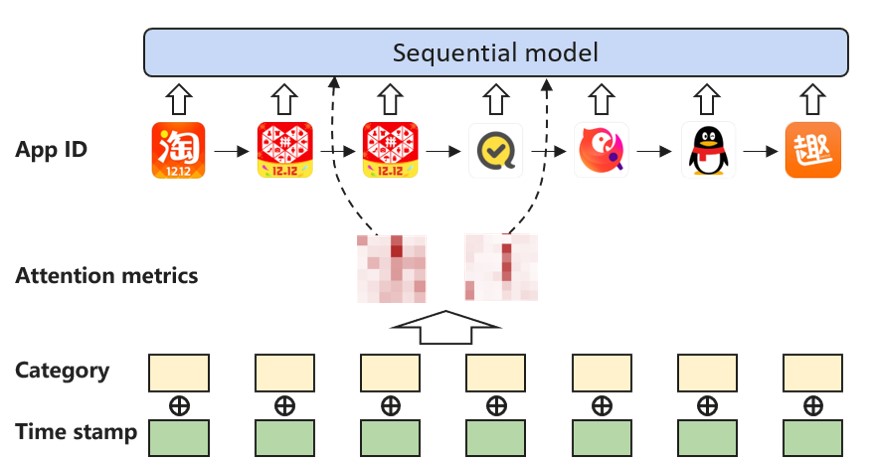}
    \caption{Non-invasive fusion approaches}
  \end{subfigure}

\caption{An illustration of invasive and non-invasive methods. The invasive methods fuse all kinds of information irreversibly, and then feed them into sequential models. For the non-invasive method, the side information only participants in the attention matrix calculation, while item information is kept in an independent vector space.}
\label{fig:intro-fig}
\end{figure}

\section{Introduction}
Recommender systems aim to model users' profiles for personalized recommendations. They are challenging to design yet business valuable. The sequential recommendation task is to predict the next item a user would be interested in, given his historical behaviors \cite{fang2019deep}. Compared with user-level or similarity-based static methods, sequential recommender systems also model the users' varying interests, hence are considered more appealing for real applications.

Recently, there is a trend to apply neural network approaches for sequential recommendation tasks, such as RNN and CNN frameworks \cite{fang2019deep}. The neural networks are now widely applied and generally perform stronger than traditional models such as \cite{davidson2010youtube}, \cite{linden2003amazon}, \cite{rendle2012bpr}, \cite{koren2009matrix}, \cite{rendle2010factorizing} and \cite{he2016fusing}.

As stated in \cite{fang2019deep}, within the family of artificial neural networks, the transformer-based models \cite{vaswani2017attention} are considered more advanced in handling sequential data, for their self-attention mechanism. Studies like \cite{bert4rec, POG, chen2019behavior} that apply transformers for sequential recommendation tasks have proved their superiority over other frameworks such as CNN and RNN \cite{hidasi2015session}. The BERT (Bidirectional Encoder Representations from Transformers) \cite{devlin2018bert} is a leading transformer model with bidirectional self-attention design. With the BERT model, \cite{bert4rec} achieves the SOTA accuracy, obviously higher than the unidirectional 
transformer methods \cite{radford2018improving, kang2018self}.

Although the BERT framework has achieved SOTA performances for many tasks, it has not been systematically studied for utilizing different types of side information. Intuitively, in addition to item IDs, it is attractive to leverage the side information, such as ratings and item descriptions, for more comprehensive depictions hence higher prediction accuracy. However, the BERT framework is originally designed to take only one type of inputs (i.e., word IDs), limiting the use of side information. Through pilot experiments, we found that existing methods usually utilize side information invasively (Figure \ref{fig:intro-fig}), but with scarce effects. Theoretically, side information should be beneficial by providing more data. Nonetheless, it is challenging to design models that can efficiently make use of the extra information. 

Hence, in this research, we study how to utilize various side information efficiently under the successful BERT framework. We propose a novel Non-inVasive self-Attention mechanism (NOVA) that can consistently improve prediction accuracy with side information and achieve state-of-the-art performances in all our experiments. As shown in Figure \ref{fig:intro-fig}, with NOVA, the side information acts as an auxiliary for the self-attention module to learn better attention distribution, instead of being fused into item representations, which might cause side effects such as information overwhelming. We verified the NOVA-BERT design on both laboratory datasets and a private dataset collected from a real application store. The results prove the superiority of the proposed method. Three of our main contributions are:
\begin{enumerate}
    \item We present the NOVA-BERT framework, which can efficiently employ various side information for sequential recommendation tasks.
    \item We also propose the non-invasive self-attention (NOVA) mechanism, a novel design that enables self-attention for compounded sequential data.
    \item Detailed experiments and deployment are conducted to prove the NOVA-BERT's effectiveness. We also include visualization analysis for better interpretability.
\end{enumerate}

\section{Related Works}
The techniques for recommender systems have evolved for a long time. Recently, researchers tend to adopt neural networks in the recommender system domain for their powerful ability \cite{fang2019deep}. Moreover, among the DNNs, the trend has also shifted from CNN to RNN, and then Transformers \cite{vaswani2017attention}. One of the Transformer-based models, BERT \cite{devlin2018bert} is considered an advanced sequential neural network, for its bidirectional self-attention mechanism. \cite{bert4rec} has proven the superiority of BERT by achieving the SOTA performances.

In the sequential recommendation domain, it is also a long-discussed topic to make full use of side information to improve accuracy. As shown in Table \ref{tab:table-intro}, under different frameworks such as CNN, RNN, Attention models, and BERT, several previous works tried to leverage side information. Nonetheless, most of them use side information without too much investigation on how side information should be added. Almost all of them use a simple kind of practices we call `invasive' fusion.

\paragraph{Invasive approaches} Most previous works, like \cite{hidasi2016parallel}, directly fuse side information into the item representations, as shown in Figure \ref{fig:intro-fig}(a). They usually use fusing operations (e.g., summation, concatenation, gated sum) to merge extra information with the item ID information, and then feed the mixture to the neural network. We call this kind of direct merging practices 'invasive approaches' because they alter the original representations.

As shown in Table \ref{tab:table-intro}, previous CNN and RNN works have tried to leverage side information by directly fusing the side information into item embeddings with operations such as concatenation and addition. Some other works like GRU \cite{lei2019tissa} and \cite{hidasi2016parallel} proposed a more complex mechanism of feature fusion gate and other training tricks, trying to make feature selection a learnable process. However, according to their experiment results, simple approaches cannot effectively leverage the rich side information under various scenarios. Although \cite{hidasi2016parallel} improves prediction accuracy by deploying a parallel subnet for each type of side information, the model becomes cumbersome and inflexible.

Another research, \cite{IDT}, does not alter the item embeddings directly, but includes dwell time with RNN models by the trick called item boosting. The general idea is to let the loss function be aware of dwell time. The longer a user looks at an item, the more interested he/she is. However, this trick heavily depends on heuristics and is limited to behavior-related side information. On the other side, some item-related side information (e.g., price) describes items' intrinsic features, which are not like dwell time and cannot be easily utilized by this approach.

\begin{table*}[ht]
\centering
\scalebox{0.70}{
\begin{tabular}{|c|c|c|c|c|}
\hline
\multicolumn{1}{|c|}{} & Addition(weighted) & Concatenation & Gating & Loss fusion  \\ \hline
CNN &  & CASER\cite{Caser} &  &  \\ \hline
RNN & p-RNN\cite{hidasi2016parallel} & MCI\cite{MCI}, Zhang.\cite{zhang2014sequential} & TISSA\cite{lei2019tissa} & IDT\cite{IDT} \\ \hline
Attention-based & SASRec\cite{kang2018self} & CSAN\cite{huang2018csan},ATrank\cite{ATRank} & TISSA\cite{lei2019tissa} &  \\ \hline
Hybrid & & M3C \& M3R \cite{M3} & & \\ \hline
BERT & BERT4Rec\cite{bert4rec} & Chen.\cite{chen2019behavior} &  &  \\ \hline
\end{tabular}}
\caption{Previous studies involving side information fusion. The first column represents different backbones, and the first row represents different fusion methods.}
\label{tab:table-intro}
\end{table*}

\section{Methodology}
In this section, we introduce the research domain and our methodology in detail. In Sec.~3.1 to 3.3, we specify the research problem and explain side information, BERT, and self-attention. Then, we present our Non-invasive Self-attention and different fusion operations in Sec.~3.4. Finally, the NOVA-BERT model is illustrated in Sec.~3.5. We denote \textbf{NO}n-in\textbf{V}asive self-\textbf{A}ttention by \textbf{NOVA} in short.

\subsection{Problem Statement}
Given a user's historical interactions with the system, the sequential recommendation task asks the next item will be interacted with, or the next action will be made. Let $u$ denotes a user, his/her historical interactions can be represented as a chronological sequence:
$$\mathbb{S}_u=[v_u^{(1)}, v_u^{(2)}, \dots, v_u^{(n)}]$$
where the term $v_u^{(j)}$ represents the $j$th interaction (also known as behavior) the user has made (e.g., download an APP). When there is only one type of actions and no side information, each interaction can be represented by simply an item ID:
$$v_u^{(j)}=ID^{(k)}$$
where $ID^{(k)} \in \mathbb{I}$, denoting the $k$th item's ID. $$\mathbb{I}=\{ID^{(1)}, ID^{(2)}, \dots, ID^{(m)}\}$$ is the vocabulary of all items to be considered in the system. $m$ is the vocabulary size, indicating the total number of items in the problem domain.

Given the history of a user $\mathbb{S}_u$, the system predicts the next item that the user will interact with most likely:
$$
\begin{aligned}
&I_{pred}=ID^{(\hat{k})}\\
&\hat{k} = \argmax_k P(v_u^{(n+1)}=ID^{(k)}|\mathbb{S}_u)
\end{aligned}
$$

\subsection{Side Information}
Side information can be anything that provides extra useful information for recommendations, which can be categorized into two types, item-related or behavior-related.

Item-related side information is intrinsic and describes the item itself, besides item IDs (e.g., price, date of production, producer). Behavior-related side information is bounded with an interaction initiated by a user, such as type of action (e.g., purchase, rate), time of execution, or the user feedback scores. The order of each interaction (i.e., position IDs in original BERT) can also be taken as a kind of behavior-related side information.

If side information is involved, an interaction becomes:
$$
    \begin{aligned}
    &v_u^{(j)}=(I^{(k)}, b_{u,j}^{(1)}, \dots, b_{u,j}^{(q)})\\
    &I^{(k)}=(ID^{(k)}, f_k^{(1)}, \dots, f_k^{(p)})
    \end{aligned}
    \label{side_info}
$$
where $b_{u,j}^{(\cdot)}$ denotes a behavior-related side information of the $j$th interaction made by user $u$. $I^{(\cdot)}$ represents an item, containing an ID and several pieces of item-related side information $f_k^{(\cdot)}$. Item-related side information is static and stores intrinsic features of each particular item. Hence the vocabulary can be rewritten as:
$$\mathbb{I}=\{I^{(1)}, I^{(2)}, \dots, I^{(m)}\}$$
The goal is still to predict the next item's ID:
$$
\begin{aligned}
&I_{pred}=ID^{(\hat{k})}\\
&\hat{k} = \argmax_{k} P(v_u^{(n+1)}=(I^{(k)}, b_1, b_2, \dots, b_q) | \mathbb{S}_u)
\end{aligned}
$$
where $b_1, b_2, \dots, b_q$ are the latent behavior-related side information, if behavior-related side information is considered. Noted that the model should still be able to predict the next item regardless of the behavior-related side information being assumptive or ignored.

\subsection{BERT and Invasive Self-attention}
\begin{figure}[ht]
  \centering
  \includegraphics[scale=0.5]{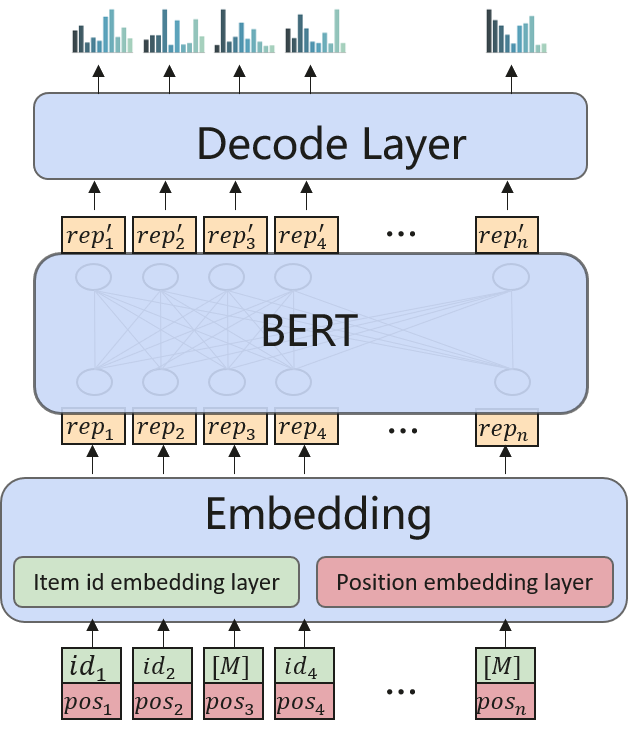}
  \caption{BERT4Rec. Item IDs and positions are encoded into vectors respectively, and then added together as integrated item representations. During training, item IDs are randomly masked (shown as $[M]$) for the model to recover.}
  \label{fig:framework}
\end{figure}

BERT4Rec \cite{bert4rec} is the first to utilize the BERT framework for sequential recommendation tasks, achieving SOTA performances. As shown in Figure \ref{fig:framework}, under the BERT framework, items are represented as vectors, which are called embeddings. During training, some items are randomly masked, and the BERT model will try to recover their vector representations and hence the item IDs, using the multi-head self-attention mechanism \cite{vaswani2017attention}:
$$
\begin{aligned}
&\textrm{SA}(Q, K, V)=\sigma(\frac{QK^T}{\sqrt{d_k}})V \\
\end{aligned}
$$
where $\sigma$ is the softmax function, $d_k$ is a scale factor, $Q$, $K$ and $V$ are derived components for query, key and value. The BERT follows an encoder-decoder design, to generate a contextual representation for each item in the input sequence. BERT employs an embedding layer to store $m$ vectors, each corresponding to an item in the vocabulary. 

To leverage side information, conventional methods like \cite{bert4rec,chen2019behavior} use separate embedding layers to encode side information into vectors, and then fuse them into the ID embeddings with a fusion function $\mathcal{F}$. This invasive kind of approach injects side information into the original embeddings, and generate a mixed representation:
$$
\label{integrated interaction embedding}
\begin{aligned}
    E_{u,j}=\mathcal{F}(&\mathcal{E}_{id}(ID), \\& \mathcal{E}_{f1}(f^{(1)}), \dots,\mathcal{E}_{fp}(f^{(p)}),
    \\& \mathcal{E}_{b1}(b_{u,j}^{(1)}), \dots, \mathcal{E}_{bq}(b_{u,j}^{(q)}))
\end{aligned}
$$
where $E_{u,j}$ is the integrated embedding for the $j$th interaction of user $u$, $\mathcal{E}$ is the embedding layer that encodes objects into vectors. The sequence of the integrated embeddings is fed into the model as the input of user history.

The BERT framework will update the representations layer by layer with self-attention mechanism:
$$
\begin{aligned}
&R_{i+1}=\textrm{BERT\_Layer}(R_i)\\
&R_1=(E_{u,1}, E_{u,2}, \dots, E_{u,n})
\end{aligned}
$$

In original BERT \cite{devlin2018bert} and Transformer \cite{vaswani2017attention}, the self-attention operation is a positional invariant function. Therefore, a position embedding is added to each item embedding to encode the position information explicitly. Position embeddings can also be viewed as a type of behavior-related side information (i.e., the order of an interaction). From this perspective, the original BERT also take positional information as the only side information, using addition as the fusion function $\mathcal{F}$.

\subsection{Non-invasive Self-attention (NOVA)}
If we consider the BERT framework end-to-end, it is an auto-encoder with stacked self-attention layers. The identical embedding map is used in both encoding item IDs and decoding the restored vector representations. Therefore, we argue that the invasive methods have the drawback of compound embedding space, because item IDs are irreversibly fused with other side information. Mixing the information from IDs and other side information might make it unnecessarily difficult for the model to decode the item IDs. 

Accordingly, we proposed a novel method called non-invasive self-attention (NOVA), to maintain the consistency of embedding space, while exploiting side information to model the sequences more efficiently. The idea is to modify the self-attention mechanism and carefully control the information source of the self-attention components, namely query Q, key K, and value V  \cite{vaswani2017attention}. More than the integrated embedding $E$ defined in Section \ref{integrated interaction embedding}, the NOVA also keeps a branch for pure ID embeddings:
$$E_{u,j}^{(ID)}=\mathcal{E}_{id}(ID)$$
Hence, for NOVA, the user history now consists of two sets of representations, the pure ID embeddings and the integrated embeddings:
$$
\begin{aligned}
&R_u^{(ID)}=(E_{u,1}^{(ID)}, E_{u,2}^{(ID)}, \dots, E_{u,n}^{(ID)})\\
&R_u=(E_{u,1}, E_{u,2}, \dots, E_{u,n})
\end{aligned}
$$
NOVA calculate $Q$ and $K$ from the integrated embeddings $R$, and V from the item ID embeddings $E^{(ID)}$. In practice we process the whole sequence in tensor form (i.e., $R$ and $R^{(ID)} \in \mathbb{R}^{B\times L\times h}$, where $B$ is the batch size, $L$ is the sequence length and $h$ is the size of embedding vectors). NOVA can be formalized as:
$$
\textrm{NOVA}(R, R^{(ID)})=\sigma(\frac{QK^T}{\sqrt{d_k}})V
$$
with $Q$, $K$, $V$ calculated by linear transform:
$$
Q=RW_Q,K=RW_K,V=R^{(ID)}W_V
$$

\begin{figure}[ht]
  \centering
  \begin{subfigure}{0.47\columnwidth}
    \includegraphics[scale=0.4]{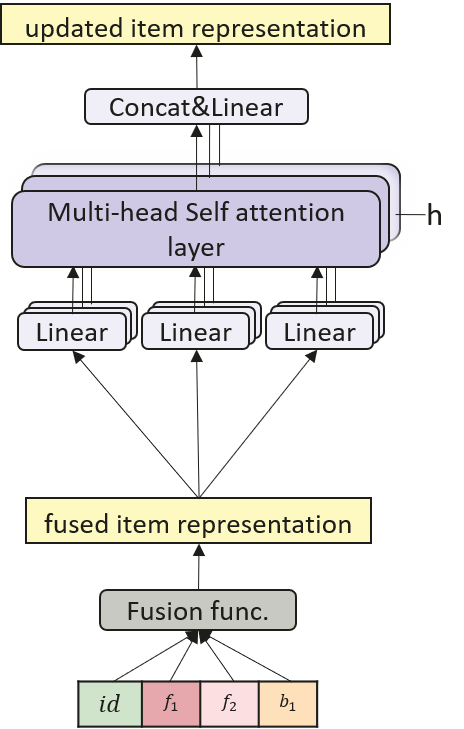}
    \caption{Invasive feature fusion}
  \end{subfigure}
  \begin{subfigure}{0.47\columnwidth}
    \includegraphics[scale=0.4]{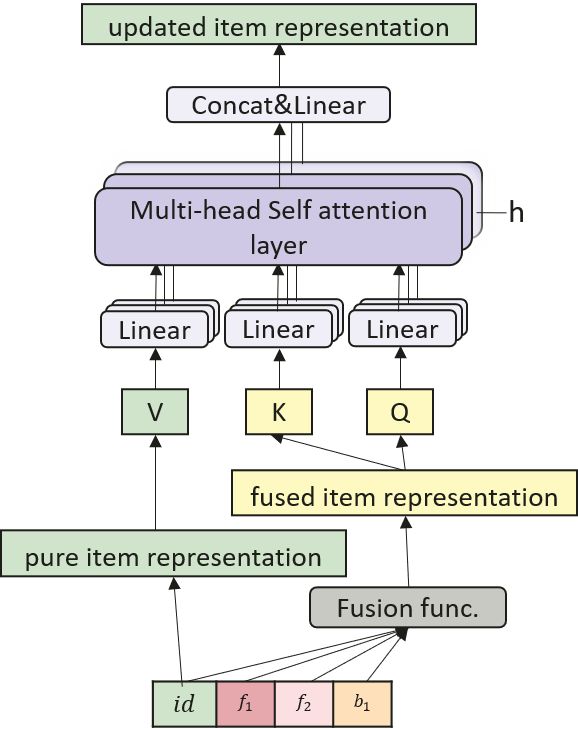}
    \caption{Noninvasive feature fusion}
  \end{subfigure}
\caption{Comparison between invasive and non-invasive ways of self-attention for feature fusion. Both fuse item-related and behavior-related side information with a fusion function, but NOVA only fuse them in the Query \& Key.}
\label{fig:Invasive_noninvasive}
\end{figure}

The comparison between NOVA and invasive ways for side information fusing is illustrated in Figure\ref{fig:Invasive_noninvasive}. Layer by layer, the representations along the NOVA layers are kept within a consistent vector space formed purely by the context of item IDs, $E^{(ID)}$.

\subsection{Fusion Operations}
NOVA leverages side information differently from invasive methods, treating it as an auxiliary and fusing side information into Keys and Queries with the fusion function $\mathcal{F}$. In this research, we also study different kinds of fusion functions and their performances.

As mentioned above, position information is also a kind of behavior-related side information, and the original BERT utilizes it with the straightforward operation of addition:
$$\mathcal{F}_{\textrm{add}}(f_1,\dots,f_m)=\sum_{i=1}^mf_i$$

Moreover, we define the `concat' fusor to concatenate all side information, followed by a fully connected layer to uniform the dimension:
$$\mathcal{F}_{\textrm{concat}}(f_1,\dots,f_m )=\boldsymbol{FC}(f_1\odot \dots \odot f_m)$$

Inspired by \cite{lei2019tissa}, we also design a gating fusor with trainable coefficients derived from the context:
$$
\begin{aligned}
&\mathcal{F}_{\textrm{gating}}(f_1,\dots,f_m )= \sum_{i=1}^mG^{(i)} f_i \\
&G=\sigma(FW^F)
\end{aligned}$$
where $F$ is the matrix form of all the features $[f_1,\dots,f_m]\in \mathbb{R}^{m\times h}$ and $W^F$ is a trainable parameter of $\mathbb{R}^{h\times1}$, $h$ is the dimension of feature vectors to be fused, $f_i \in \mathbb{R}^h$

\subsection{NOVA-BERT}
\begin{figure}[ht]
  \centering
  \includegraphics[scale=0.5]{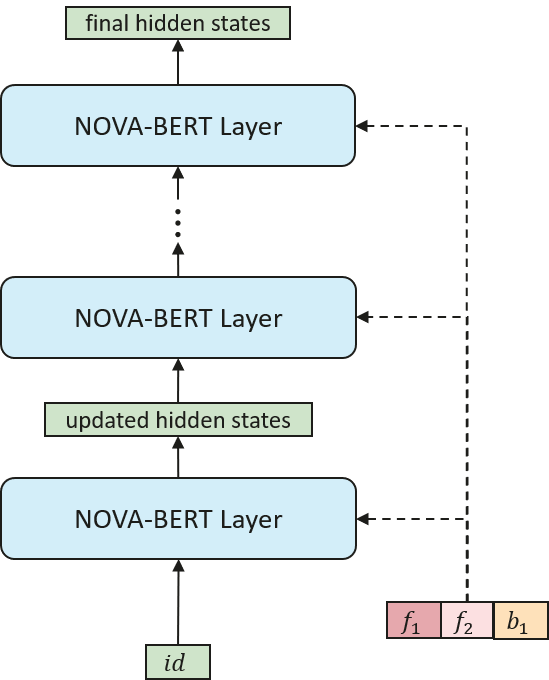}
  \caption{NOVA-BERT. Each NOVA layer takes two inputs, item representations and side information.}
  \label{fig:noninvasiveBERT}
\end{figure}

As illustrated in Figure \ref{fig:noninvasiveBERT}, we implement our NOVA-BERT model under the BERT framework with the proposed NOVA operation. Each NOVA layer takes two inputs, the supplementary side information and the sequences of item representations, then outputs updated representations of the same shape, which will be fed to the next layer. For the input of the first layer, the item representations are pure item ID embeddings. Since we only use side information as auxiliary to learn better attention distributions, the side information does not propagate along with NOVA layers. The identical set of side information is explicitly provided for each NOVA layer.

The NOVA-BERT follows the architecture of the original BERT in \cite{devlin2018bert}, except replacing the self-attention layers by NOVA layers. Hence, the extra parameters and computation overheads are negligible, mainly introduced by the lightweight fusion function.

We believe that with NOVA-BERT, the hidden representations are kept in the same embedding space, which will make the decoding process a homogeneous vector search and benefit the prediction. Results in the next section also empirically verify the effectiveness of NOVA-BERT.

\section{Experiments}
In this section, we evaluate the NOVA-BERT framework on both public and private datasets, aiming to discuss NOVA-BERT more in the following aspects:
\\
\textbf{1. Effectiveness}. Whether the NOVA-BERT approach outperforms conventional invasive feature fusion methods?
\\
\textbf{2. Thoroughness}. How do the components of NOVA-BERT  contribute to the improvements (e.g., different types of side information, fusion function)?
\\
\textbf{3. Interpretability}. How does side information affect the attention distribution hence improve prediction accuracy?
\\
\textbf{4. Efficiency}. How does NOVA-BERT perform in terms of efficiency for future real-world deployments?

\subsection{Experiment Settings}
\begin{table}[ht]
\centering
\fontsize{9}{11}\selectfont 
\setlength{\tabcolsep}{0.2em}
\begin{tabular}{ccccc}
\hline
Dataset & \#records & \#users & Len & Side information available                                                       \\ \hline
ML-1m   & 988,129    & 6,040    & 165         & year, genres, rating          \\
ML-20m  & 19,723,277  & 138,493  & 142         & year, genres, rating          \\
APP     & 1,774,867   & 100,000  & 18          & category, developer, size, tag \\ \hline
\end{tabular}
\caption{Dataset statistics. `Len' for mean sequence length.}
\label{tab:dataStatistic}
\end{table}

\paragraph{Datasets}
We evaluated our methods on the public MovieLens datasets and a real-world dataset called APP. Please refer to Table \ref{tab:dataStatistic} for detailed descriptions of the datasets.

We sort the rating/downloading records by chronological order, to construct each user's interaction history. Following the practice of \cite{kang2018self,bert4rec}, we use the heading subsequence for each user's record without the last two elements as the training set. The second last items in the sequences are used as the validation set for tuning hyper-parameters and finding the best checkpoint. The last elements from these sequences construct the test set. Short sequences having less than five elements are discarded to eliminate disturbances from the cold-start problem. The experiment aims to focus on side information fusion and hold a fair comparison with other methods. 

\paragraph{Hyper-parameter Settings}
For all the models in this research, we train them with Adam optimizer using a learning rate of 1e-4 for 200 epochs, with a batch size of 128. We fix the random seeds to alleviate the variation caused by randomness. The learning rate is adjusted by a linear decay scheduler with a 5\% linear warm-up. 

We also apply grid-search to minimize the bias of our experiment results. The searching space contains three hyper-parameters, $hidden\_size\in \{128,256,512\}$, $num\_heads\in \{4,8\}$ and $num\_layers\in \{1,2,3,4\}$. In the end we use 4 heads and 3 layers, as well as 512 hidden size for MovieLens Datasets and 256 for the APP dataset.

\begin{table*}[ht]
\centering
\fontsize{9}{11}\selectfont 

\setlength{\tabcolsep}{0.2em}
\begin{tabular}{c|c|cccc|cccc|cccc}
\hline
\multirow{2}{*}{Method} & \multirow{2}{*}{Fusion func} & \multicolumn{4}{c|}{ML-1m} & \multicolumn{4}{c|}{ML-20m} & \multicolumn{4}{|c}{APP} \\ \cline{3-14} 
 &  & HR@10 & ND@10 & HR@5 & ND@5 & HR@10 & ND@10 & HR@5 & ND@5 & HR@10 & ND@10 & HR@5 & ND@5 \\ \hline
BERT4Rec & Add (posID) & 25.24 & 13.98 & 16.61 & 11.20 & 30.70 & 18.83 & 22.89 & 16.31
 & 27.44 & 16.39 & 19.73 & 13.91 \\ \hline
\multirow{3}{2.4cm}{BERT + Invasive methods} & Add & 24.65 & 13.73 & 16.82 & 11.23 & 31.00 & 18.94 & 23.09
 & 16.39 & 27.57 & 16.45 & 19.82 & 13.95 \\
 & Concat & 24.92 & 14.00 & 16.57 & 11.30 & 31.21 & 19.21 & 23.38 & 16.68 & 27.38 & 16.11 & 19.38 & 13.53 \\
 & Gating & 24.60 & 13.61 & 16.39 & 10.98 & 31.29 & 19.25 & 23.44 & 16.71 & 27.61 & 16.36 & 19.72 & 13.82 \\ \hline
\multirow{3}{*}{\begin{tabular}[c]{@{}l@{}}NOVA-BERT\end{tabular}} & Add & 28.34 & 16.55 & 20.22 & 13.95 & 31.65 & 19.40 & \textbf{23.59} & 16.76 & 27.73 & 16.54 & 19.88 & 14.02 \\
 & Concat & 27.70 & 16.18 & 19.82 & 13.64 & \textbf{31.68} & \textbf{19.42} & \textbf{23.59} & \textbf{16.81} & \textbf{28.00} & 16.65 & 20.02 & 14.08 \\
 & Gating & \textbf{28.65} & \textbf{16.80} & \textbf{20.45} & \textbf{14.16} & 31.22 & 19.15 & 23.30 & 16.60 & 27.85 & \textbf{16.68} & \textbf{20.08} & \textbf{14.17} \\ \hline
\end{tabular}

\caption{Experiment results in \%, `ND' stands for the NDCG metric. We evaluate NOVA-BERT by comparing it with the SOTA model BERT4Rec~\cite{bert4rec} and its invasive modifications. The best results are boldfaced. For each dataset, we use all available side information. The baseline BERT4Rec, however, does not support other side information except position ID.}
\label{tab:Performance}
\end{table*}

\subsection{Evaluation Metrics}
Following \cite{kang2018self}, we employ two widely-adopted evaluation metrics: Hit Ratio(HR@k), and Normalized Discounted Cumulative Gain (NDCG@k), with $k\in \{5,10\}$. Particularly, we rank the whole vocabulary, which means the model gives a prediction score on every item, indicating the degree of recommendation.

Some practices, such as \cite{bert4rec} uses a smaller candidate set, for example, the ground truth plus 100 randomly sampled negative items with chance proportional to popularity. During pilot experiments, we found that the practice may result in severe bias. For example, the unpopular ground-truth item is obvious among other negative candidates, who are drawn because of their high popularity. The model tends to take advantage of the loose metric. Therefore, we rerun the baselines with the ranking-all metric.

\subsection{Experiment Results}
As stated before, \cite{bert4rec} has proven the extraordinary performance of the BERT framework under a wide range of scenarios. Therefore, in this paper, we choose their SOTA method as the baseline and focus on leveraging side information under the BERT framework. We also examine the fusion functions of addition, concatenation, and gating. 

In Table \ref{tab:Performance}, the NOVA-BERT outperforms all other methods on the three datasets and all metrics. Compared with the Bert4Rec\cite{bert4rec} exploiting only position IDs, the invasive approaches use several kinds of side information but have very limited or even adverse improvements. On the contrary, the NOVA-BERT can effectively leverage the side information and stably outperform all other methods. 

However, the improvements brought by NOVA are on different scales for the datasets. In our experiments, for larger and denser datasets, the scale of improvement decreases. For the ML-1m task, NOVA with gating fuser outperforms the baseline by 13.51\% in HR@10, while all invasive methods perform worse than the baseline. However, in ML-20M and APP tasks, the relative improvements brought by noninvasive approaches decrease to 3.19\% and 1.48\%, respectively. We hypothesize that with a more affluent corpus, the models are possible to learn good enough item embeddings even from the item context itself, leaving a smaller space for side information to make supplements.

Furthermore, the results also demonstrate the robustness of NOVA-BERT. No matter what fusion function is employed, the NOVA-BERT can consistently outperform the baselines. The best fusion function may depend on the dataset. In general, the gating method has a strong performance, possibly benefited from its trainable gating mechanism. The results also suggest that for real-world deployments, the type of fusion function can be a hyper-parameter and should be tuned according to the dataset's intrinsic properties, to reach the best online performance.

\begin{figure*}[ht]
  \centering
  \begin{subfigure}{0.48\textwidth}
    \centering
    \includegraphics[scale=0.15]{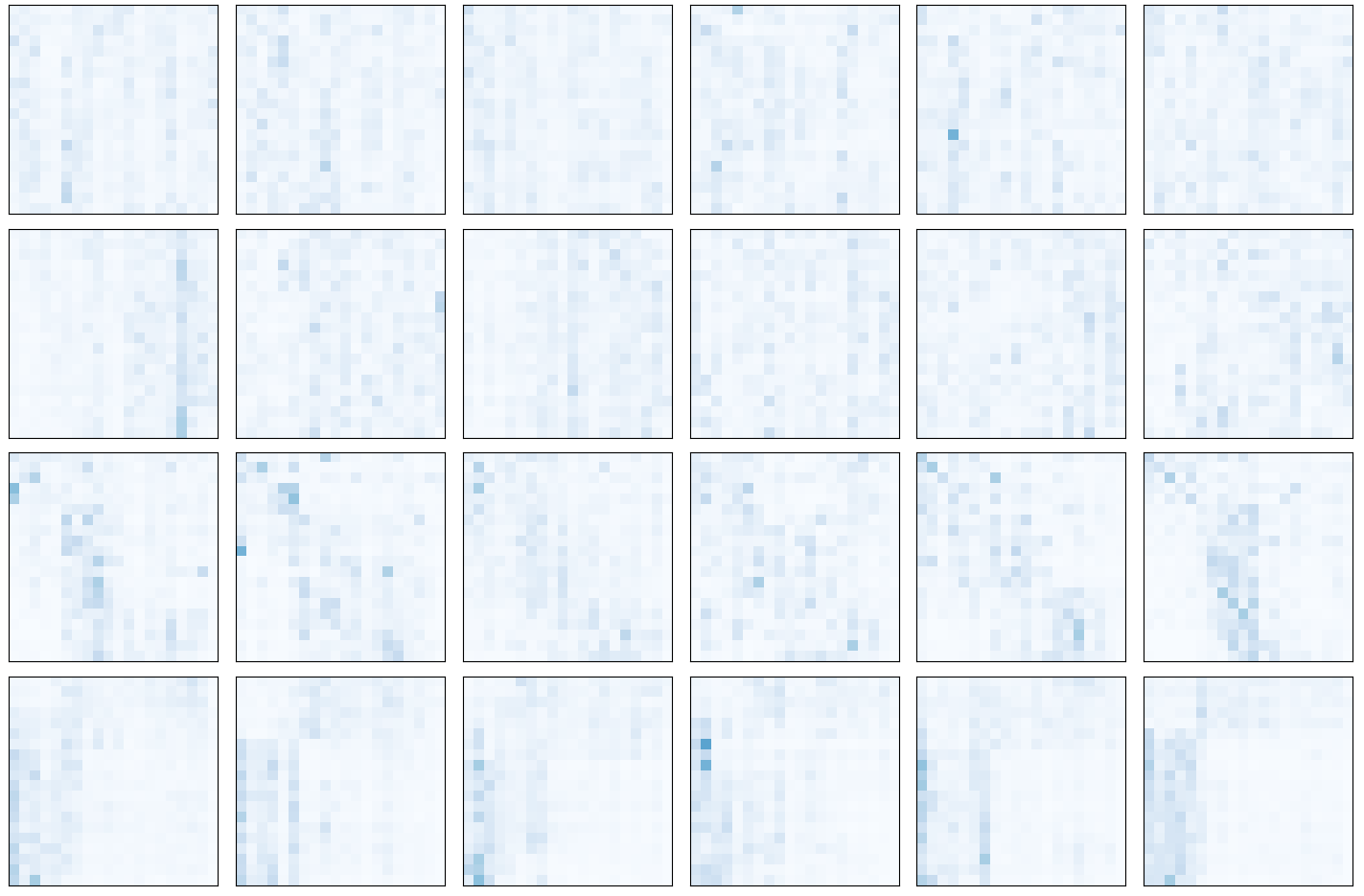}
    \caption{BERT4Rec. Position ID is the only side information supported.}
  \end{subfigure}\hfill
  \begin{subfigure}{0.48\textwidth}
    \centering
    \includegraphics[scale=0.15]{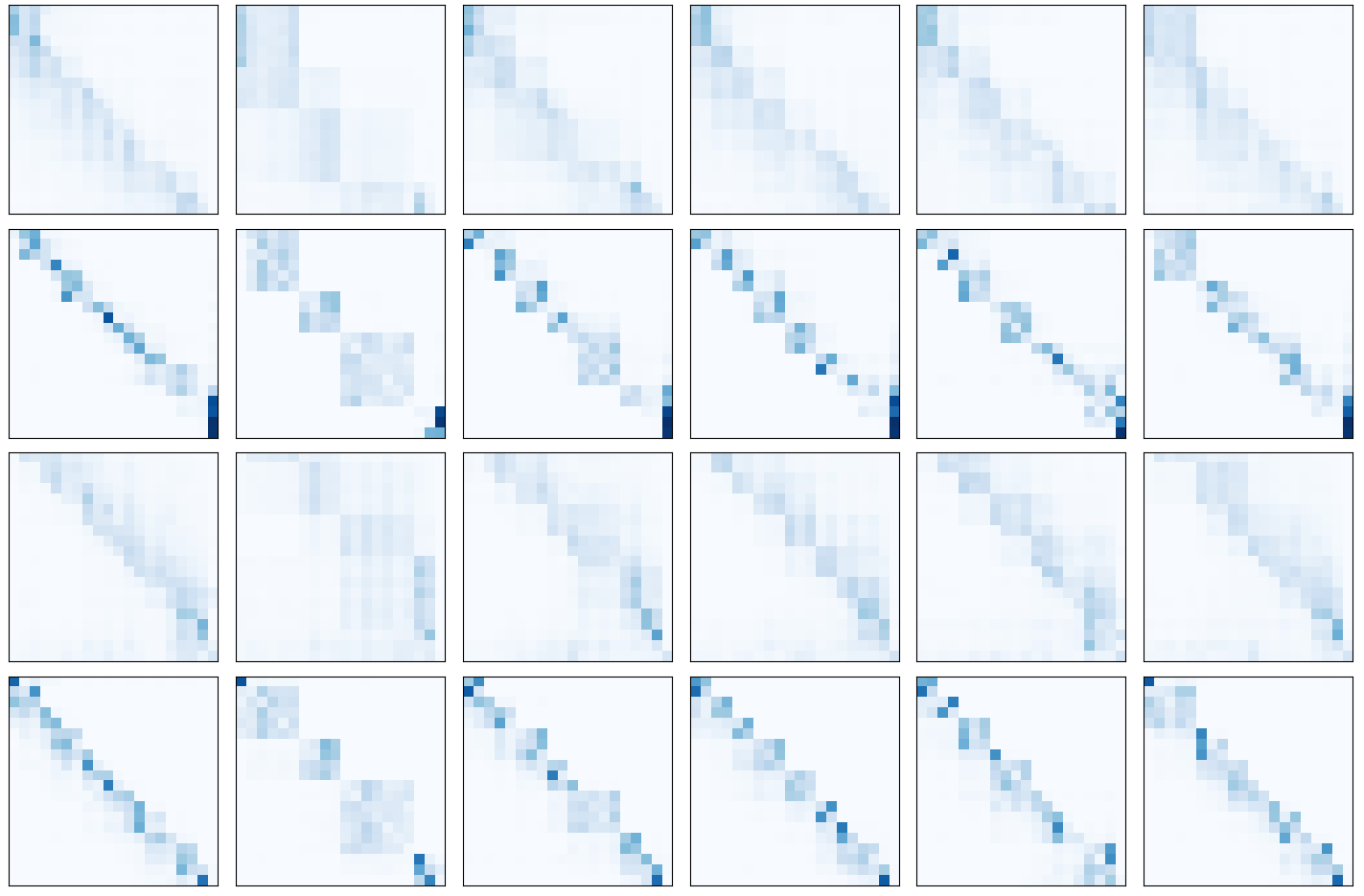}
    \caption{NOVA-BERT. All available side information employed.}
  \end{subfigure}

\caption{Influence of side information on attention distribution. The graphs show attention matrices of different attention heads (4 rows) from random samples (6 columns). Each matrix denotes the item-to-item attention. The darker the color, the higher the score. Compared with the original BERT4Rec, the NOVA-BERT presents an articulate pattern of attention distribution.}
\label{fig:InfluenceSideInfo}
\end{figure*}

\subsection{Contributions of Different Side Information}
In Section \ref{side_info}, we categorized side information as item-related and behavior-related. We also investigate the contributions of the two types of side information under the \textbf{ML-1m} task. By definition, publishing year and category are item-related, while rating score is behavior-related.

Table \ref{tab:sideInfoAblation} shows the results of different types of side information. The original BERT framework considers no side information except the position ID, denoted as \textit{None} in the first row. When complete side information is given, the NOVA-BERT presents the most significant improvement. Meanwhile, in separate, the item-related and behavior-related side information bring nonobvious benefits to the accuracy. Since item-related side information (year and genre) is intrinsic features of the movies, it might be latently derived from the massive data sequences, hence resulting in the similar accuracy. On the other hand, if behavior-related side information is also combined, the improvement is noticeable greater than the sum of improvements brought by either of them. This observation suggests that the impacts brought by different types of side information are not independent. Moreover, the NOVA-BERT benefits more from synthesis side information and has a strong ability to utilize the rich data without being troubled by information overwhelming.

In the case of \textbf{ML-1m}, comprehensive side information can boost the accuracy most under NOVA-BERT. However, it is reasonable to believe that the contributions of different types of side information also depends on the data quality. For that reason, we also consider side information selection as a dataset dependent factor. 

\begin{table}[ht]
\centering
\fontsize{9}{11}\selectfont 
\setlength{\tabcolsep}{0.4em}
\begin{tabular}{lcccc}\hline
Side-info & HR@10 & NDCG@10 & HR@5 & NDCG@5 \\ \hline
None & 0.2809 & 0.1636 & 0.2015 & 0.1383 \\
Item & 0.2743 & 0.1609 & 0.1927 & 0.1346 \\
Behavior & 0.2815 & 0.1655 & 0.2005 & 0.1393 \\
All & 0.2865 & 0.1680 & 0.2045 & 0.1416 \\ \hline
\end{tabular}
\caption{Ablation study on side information type. NOVA-BERT with gating fuser is tested on the ML-1m dataset. Side information of each row: \Romannum{1}) no extra side information (posID only), \Romannum{2}) item-related (year+genre), \Romannum{3}) behavior-related (rating), \Romannum{4}) all avaliable side information.}
\label{tab:sideInfoAblation}
\end{table}

\subsection{Visualization of Attention Distribution}
To provide more discussions on NOVA-BERT's interpretability, we visualize the attention distribution of the bottom NOVA layer. In Figure \ref{fig:InfluenceSideInfo}, we present the comparison of attention scores of 6 randomly chosen samples (columns), each row represents a head from the multi-head self-attention. Every pixel denotes the attention intensity from the row item to the column item. All the attention scores in a row sum to 1. In Figure \ref{fig:InfluenceSideInfo}(a), the attention scores from the original BERT4Rec model is visualized. In Figure \ref{fig:InfluenceSideInfo}(b), we show the attention scores of the NOVA-BERT. The model is fed with all available kinds of side information.

As shown in the graph, the attention scores from the NOVA-BERT show a stronger pattern in terms of locality, concentrating approximately along the diagonal. On the other side, it is not observed in the graph for the baseline model. The contrast is prevailing, according to our observation over the whole dataset. We noticed that side information leads the model to form more explicit attention in early layers. The observation demonstrates that the NOVA-BERT, which takes side information as an auxiliary for calculating attention matrix, can learn targeted attention distribution and hence improve the accuracy.

Additionally, we also conduct case studies to understand more about the attention distribution of NOVA, with more detailed analysis. Please refer to the Appendix for details.

\subsubsection{More explanations on NOVA's attention matrix}
Because of the sophisticated high dimension embedding space and the intricate network architecture, it is hard to fully comprehend how and why an item attended the other. However, from the visualization results, we found that with side information provided, there is a trend for the NOVA-BERT to learn a more structured pattern of attention scores.

Due to the limited interpretability of deep neural networks, we cannot assert that NOVA-BERT's attention pattern is absolutely better. Nonetheless, we found the observed patterns similar to those in the study \cite{kovaleva2019revealing}, which presents a correlation between this kind of attention distributivity and better performances in natural language processing tasks. Additionally, NOVA-BERT also shows the correlation in sequential recommendation tasks, suggesting that attention is indeed smartly distributed.

\section{Deployment and Cost of NOVA-BERT}
NOVA-BERT consistently surpasses the current deployed methods in real-world scenarios, according to our online tests. In Table~\ref{tab:model_cost}, the efficiency results of NOVA-BERT and part of the baseline models are listed. We evaluate the models in terms of FLOPs and sizes, with `addition' being the fusion function. As shown in the table, NOVA-BERT has almost no extra computational overheads and the same size as the invasive method. Moreover, since the NOVA-BERT is well supported by parallel computing and GPU acceleration, the inference time is also approximately the same as the original BERT. We also claim that since the parameters are mainly brought by extra embedding layers, for deployment we may use look-up table to replace the embedding layers and result in similar model size as the original BERT.

\begin{table}[ht]
\centering
\fontsize{9}{11}\selectfont 
\begin{tabular}{lcc}\hline
  & \#FLOPs ($\times 10^9$)& Size (MB)\\ \hline
Ori-BERT & 0.4796 & 35.5  \\
Invasive-add & 0.4986 & 42.6  \\
NOVA-BERT (add) & 0.4988 & 42.6  \\ \hline
\end{tabular}
\caption{Complexity in computation and model size.}
\label{tab:model_cost}
\end{table}

\section{Conclusion and Future Work}
In this work, we present the NOVA-BERT recommender system and the non-invasive self-attention mechanism (NOVA). Instead of fusing side information directly into the item representations, the proposed NOVA mechanism utilizes the side information as directional guidance and keeps the item representations undoped in their vector space. We evaluate the NOVA-BERT on both experimental datasets and industrial applications, achieving SOTA performances with negligible overheads in computation and model size.

Although the proposed method reaches the SOTA performance, there are still several intriguing directions for future studies. For example, fusing side information at every layer may not be the best approach, and stronger fusing functions are also expected. Additionally, we will continue studying and evolving our approach for higher online performances as well as deploy it in industrial products.

\bibliography{RecBERT}

\end{document}